\documentclass{aa}
\usepackage{graphicx,natbib}
\bibpunct{(}{)}{;}{a}{}{,}

\def\apj{ApJ}

\def\flux{erg s$^{-1}$ cm$^{-2}$}
\def\lum{erg s$^{-1}$}

\begin{document}

\sloppypar

%
   \title{On the contribution of point sources to the Galactic ridge
X-ray emission}

   \author{M.~Revnivtsev \inst{1,2} \and S.~Sazonov \inst{1,2}}

   \offprints{mikej@mpa-garching.mpg.de}

   \institute{
              Max-Planck-Institute f\"ur Astrophysik,
              Karl-Schwarzschild-Str. 1, D-85740 Garching bei M\"unchen,
              Germany,
      \and
              Space Research Institute, Russian Academy of Sciences,
              Profsoyuznaya 84/32, 117997 Moscow, Russia
            }
  \date{}

        \authorrunning{Revnivtsev et al.}
        \titlerunning{Weak point sources in the Galactic plane region}

   \abstract{We analyzed deep {\it Chandra} observations of the Galactic 
plane region centered at $l=28.55^\circ$, $b=-0.03^\circ$ with the
aim to obtain the best possible constraints on the contribution of weak
point sources to the Galactic ridge X-ray emission (GRXE) in this
region. We demonstrate that the vast majority of the detected sources
are Galactic in origin and are probably cataclysmic variables and
coronally active stars. We use the number-flux function of detected
sources to constrain the luminosity function of Galactic X-ray sources
in the range $10^{30}$--$10^{32}$ \lum\ and find good agreement with
the luminosity function of sources in the Solar vicinity. The fraction
of the total flux at energies 1--7~keV resolved into point sources at
the current sensitivity level is $\sim 25$\%. Excluding the expected
contribution of extragalactic sources, $\sim 19$\% of the GRXE is due
to point Galactic sources with interstellar absorption
corrected fluxes higher than $1.2\times 10^{-15}$ \flux in the energy band
1-7 keV. 
   }

   \maketitle

%

\section{Introduction}

The origin of the Galactic ridge X-ray emission (GRXE) has been a
long-standing problem of X-ray astronomy. Since the discovery of this
emission in the late 1970s \citep{cooke70,bleach72,worrall82},
different explanations have been proposed for its nature. These can be
divided into two major branches: i) the GRXE is truly diffuse, i.e. it
is generated in the interstellar medium \cite[see
e.g.][]{koyama86,koyama89,tanaka02,ebisawa05}, or ii) it is the
integrated emission of faint Galactic point sources
\citep{worrall83,koyama86,ottmann92,mukai93,mikej06,mikej06_67kev}.

The main problem with the former hypothesis is the difficulty of keeping
plasma that is apparently thermal and hot ($>$5--10 keV) 
\citep{koyama86,koyama89,tanaka02,koyama06} within a thin layer along 
the Galactic plane \cite[see e.g.][]{worrall82,warwick85,yamauchi93,mikej06}.

The latter scenario recently received strong support from work of our
team \citep{mikej06,sazonov06,krivonos06,mikej06_67kev}. In
particular, we obtained high-quality maps of the GRXE that showed striking 
similarity with the near-infrared map representing the distribution of
stellar mass over the Milky Way. We concluded that throughout the
Galaxy the GRXE volume emissivity is proportional to the stellar 
mass density. Furthemore, we showed that the GRXE emissivity
per unit stellar mass is consistent with the cumulative emissivity of
known classes of weak X-ray sources in the Solar vicinity -- 
cataclysmic variables and coronally active stars \citep{sazonov06}. 

We recently used deep {\em Chandra} observations of a field in the
close vicinity of the Galactic Center (GC) to demonstrate that at
least 40--50\% of the total X-ray emission from that region is created
by point sources \citep{mikej_gc}. It is known that the properties of the X-ray
emission from the GC region are very similar to those of the
large-scale GRXE, suggesting that both have the same physical 
origin \cite[e.g.][]{tanaka02}. However, as the space
density of stars near the GC is 3--4 orders of magnitude higher than
is typical for the Galactic plane (GP), one could expect that truly
diffuse X-ray emission, should such exist, might provide a more 
significant contribution to the GRXE from the GP.

This motivated us to study one of the deepest {\em Chandra}
observations of the GP, centered at $l=28.55^\circ$,
$b=-0.03^\circ$, with the aim to estimate the contribution of resolved
Galactic point sources to the total X-ray flux from this
direction. This observation was previously analyzed by 
\cite{ebisawa05}. 

\section{Data reduction}
\label{sect:data}
We used the {\em Chandra} observation (OBSID 2298) of the GP region in
the direction $l=28.5548^\circ, b=-0.026^\circ$, with a 
total exposure time of $\sim 102$ ksec. In order to maximize the 
sensitivity to point sources, we restricted our analysis to the data
collected within $4\arcmin$ radius around the aim point of the
telescope, where its angular resolution is better than 
$1''$ (FWHM). The solid angle of the field of our study is thus
$\sim4.25 \times 10^{-6}$~ster. Note that we did not analyze existing
{\em Chandra} observations of a nearby GP region (OBSIDs 949 and 1523),
because the central $4\arcmin$-area of this field does not overlap
with the region of our study, making it impossible to increase the
point source sensitivity, which is key to us, by adding these extra data. 

The data were reduced following a standard procedure fully described in
\cite{2005ApJ...628..655V}. The only difference is that the detector
background was now modeled using the stowed dataset
(http://cxc.harvard.edu/contrib/maxim/stowed). Point sources were
detected using the wavelet decomposition package $wvdecomp$ of
$ZHTOOLS$ software
\citep{vikhlinin98}\footnote{http://hea-www.harvard.edu/saord/zhtools/}.

Due to the significant drop in the {\it Chandra} sensitivity at energies 
$E>2$ keV and especially at $E>7$~keV, and because of the strong
effect of interstellar photoabsorption in the GP at $E<1$ keV, we used  the
energy band 1--7 keV for our analysis. A total of 6.5 kcnts
in this energy band were accumulated from the studied area of 0.01396
sq. deg over 102 ksec. This implies the total observed X-ray 
flux $F_{\rm 1-7~keV,~observed}= (1.0\pm0.1)\times 10^{-12}$
\flux. 

After constructing the number-flux function of deteced sources, we
corrected it for the incompleteness near the 
detection threshold using the procedure developed for our GC study
\citep{mikej_gc}. In essence, Poisson fluctuations in the number of
counts detected from a weak source effectively reduce the probability
of its detection in a given observation \cite[see
e.g.][]{hasinger93,kenter03}. To correct for this ``leakage'' of
sources under the detection threshold, we derived the allowed range of
{\sl intrinsic} log $N$--log $S$ functions of sources by simulating
mock {\it Chandra} images for a large number of such functions and
checking which of these images are consistent with the {\sl measured}
number-flux relation. The multifold of allowed intrinsic log $N$--log $S$
functions was then used to determine the resolved fraction of the GRXE. 

The studied {\sl Chandra} field is characterized by heavy interstellar
absorption. The total column density of atomic and molecular gas through the 
Galactic disk in this direction is estimated at $\sim6\times10^{22}$ 
H atoms per cm$^2$ \citep{ebisawa05}. The Galactic sources in the field
are expected to be located at widely differing distances, from our
close vicinity all the way to the outer boundary of the Milky Way, whereas the
distribution of the absorbing gas within the sampled cone of the
Galaxy is poorly known. \cite{ebisawa05} made an attempt to estimate
typical column densities from stacked {\sl Chandra} spectra of subsets
of sources detected in this field and found values ranging from
$\sim0.5\times10^{22}$ to $>5\times10^{22}$ cm$^{-2}$. In view of the
remaining uncertainty, we assume in our analysis that all 
Galactic X-ray sources in the field are characterized by the same
absorbing column density $N_{\rm H}=2\times10^{22}$ cm$^{-2}$, while
all extragalactic ones by $N_{\rm H}=6\times10^{22}$ cm$^{-2}$.

To make a direct comparison of the source counts in the GP
region with those in typical extragalactic fields (maximally devoid of
Galactic sources), we also analyzed {\sl Chandra} observations of the Deep
Fields South (CDFS, OBSID 5021) and North (CDFN, OBSID 3389) in
exactly the same manner as the GP data. To have the same exposure time
as for the GP field, we used only $\sim 100$ ksec of the available
(ultra-deep) observations for each of these extragalactic
fields. However, even in this case the number of detected
extragalactic sources in the GP region is expected to be significatly
reduced compared to the CDFS and CDFN by the strong interstellar absorption.

\section{Expected number-flux functions of Galactic and extragalactic sources}

\begin{figure}
\includegraphics[width=\columnwidth,bb=43 185 576 591,clip]{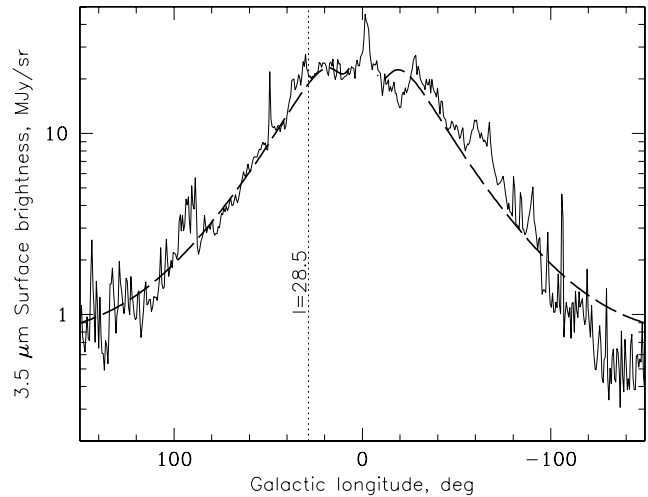}
\caption{Profile of the near-infrared brightness of the Galaxy along
its plane as seen by {\sl COBE/DIRBE} at $3.5 \mu$m. The dashed line is a
profile predicted by the Galaxy mass model adopted in this paper.}
\label{slice_equator}
\end{figure}

The majority of weak X-ray sources ($L_{\rm x}<10^{34}$ \lum) in our Galaxy
are cataclysmic variables and coronally active binary stars
\citep[e.g.][]{vaiana81,sazonov06}. These classes of sources are expected to be
distributed over the Milky Way just like ordinary stars (with the
notable exception of globular clusters where the relative fraction of X-ray
sources may be different due to the greatly increased role of
dynamical processes in forming such systems). Such proportionality
between the space density of weak X-ray sources and that of stars has
indeed been observed (see e.g. \citealt{muno06,mikej_gc}). This allows
us to predict the distribution of X-ray sources along our 
studied light of sight and consequently the expected log $N$--log $S$
function for a given source luminosity function, e.g. for 
that measured in the Solar vicinity \citep{sazonov06}.

In estimating the contribution of extragalactic sources to the 
number-flux function measured in the GP field, we preferred to carry
out a direct comparison with $\sim 100$~ksec subsets of CDFN and CDFS
observations, rather than relying on published log $N$--log $S$
functions of extaragalactic sources \cite[e.g.][]{moretti02}. This
allows us to use the same energy band (1--7~keV) for detecting sources, thus
avoiding any rescaling and its associated uncertainties. Also,
number-flux relations constructed from these observations should be
affected by the leakage of weak sources (see above) similarly to 
the GP observation, faciliating comparison of the measured log
$N$--log $S$ curves.

\subsection{Galactic disk model}
\label{disk_model}

In order to make a prediction for the number-flux function of Galactic
X-ray sources in the studied GP region, we should take into account
the spatial distribution of stars (which have been shown to be good tracers of
weak X-ray sources) along the line of sight. For this purpose we
should adopt some mass model of the Galaxy. 
Since our observation is directed relatively far from the GC
($l\approx28.5^\circ$), we do not expect any significant contribution 
from sources residing in the Galactic bulge. Thus, it is sufficient to
consider only the disk component of the Galaxy. 

There is general agreement that the space density of stars in the
Galactic disk is exponentially declining (with scale-height $R_{\rm
disk}$) with Galactocentric radius $R$ and that there is a similar
decline of density with radius within some minimal radius $R_{\rm m}$
due to the presence of the Galactic bulge/bar inside this radius
\cite[e.g.][]{freudenreich98}. We therefore adopt below the following
mass model of the Galactic disk:

\[
\rho_{\rm disk}(R)=\rho_{0, \rm disk} \,\, \exp\left[-\left({R_{\rm
m}\over{R}}\right)^3-{R\over{R_{\rm disk}}}\right],
\label{eq:disk_model}
\]
where $\rho_{0,\rm disk}=5.5 M_\odot$ pc$^{-3}$, $R_{\rm m}=2.5$ kpc,
and $R_{\rm disk}=2.2$ kpc. We additionally assumed that 
the stellar space density abruptly drops outside a Galactocentric
radius of 10 kpc and that the distance between the Sun and the GC is
8.0 kpc. As our observation is aimed almost exactly at the Galactic
plane and the angular size of the studied region is tiny
($\sim4^\prime$), it is appropriate to ignore the vertical
distribution of stars in the Milky Way.

\begin{figure*}
\includegraphics[height=\textwidth,angle=-90]{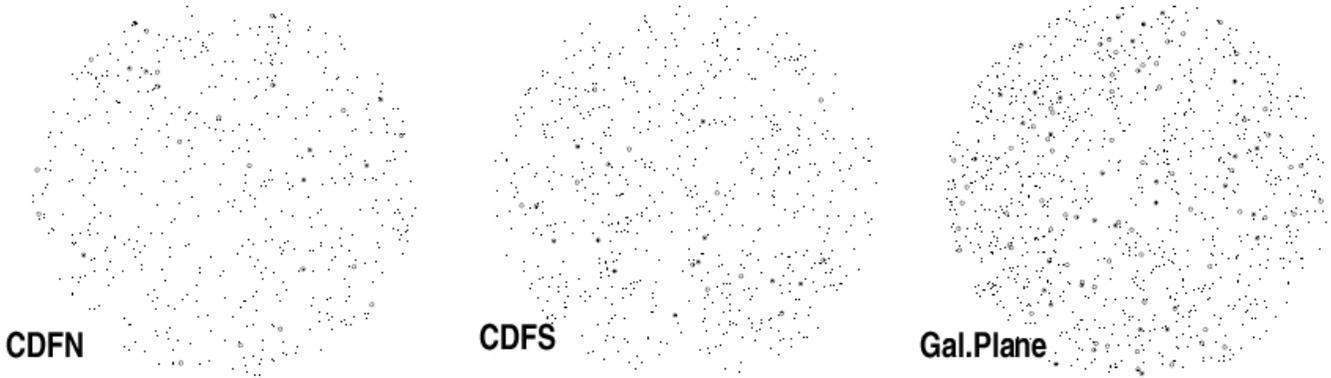}
\caption{Comparison of the raw {\sl Chandra} images (1--7~keV) of
three sky regions. From left to right -- {\sl Chandra} Deep Field
North (CDFN), {\sl Chandra} Deep Field South (CDFS), and Galactic
plane region at $l\sim 28.5^\circ$. For each of the CDFN and CDFS only
$\sim$100 ksec of the available observations are used and the same
circular region around the telescope's aim point is chosen as for the
GP field. Even with this equal setup for all three images, {\it
Chandra}'s sensitivity to extragalactic sources is expected to be much
worse in the GP region compared to both extragalactic fields. In
reality, the surface density of point sources is much higher in the GP
region, clearly indicating the presence of a large number of Galactic
sources in the field.}
\label{three_images}
\end{figure*}

The above model predicts a near-infrared (NIR) surface brightness profile of
the Galaxy along its plane as shown in Fig.~\ref{slice_equator}. For
comparison shown is the NIR profile of the Galaxy measured by {\sl
COBE/DIRBE} (corrected for the interstellar extinction following  
\citealt{mikej06_67kev}). One
can see that the predicted and measured profiles are in 
satisfactory agreement with each other. Most importantly, the model by
design correctly reproduces the observed NIR intensity in the direction of
our study ($l\sim28.5^\circ$).

\subsection{Cumulative characteristics at $l=28.5^\circ$}

The extinction corrected surface brightness of the
Galactic disk at 3.5 $\mu$m in the direction of our observation
($l\sim28.5^\circ$,  $b\sim0.03^\circ$) is $I_{\rm 3.5 \mu m}\sim
21\pm4$ MJy/sr (a significant fraction of the quoted uncertainty is associated
with the correction of the {\sl COBE/DIRBE} data for interstellar
extinction), or\footnote{We adopted the {\sl
DIRBE} bandwidths from
http://lambda.gsfc.nasa.gov/product/cobe/about\_dirbe.cfm.}
$(1.4\pm0.3) \times10^{-6}$ \flux\ deg$^{-2}$.  
Given the area ($1.39\times 10^{-2}$ deg$^{-2}$) of the studied {\sl
Chandra} field, the total NIR flux from this region $F_{\rm 3.5 \mu
m}=(2.0\pm0.4)\times10^{-8}$ \flux.

From our study of the large-scale distribution of the GRXE with {\sl RXTE} 
we found the ratio of the NIR surface brightness to the GRXE intensity
in the 3--20~keV band to be $(4.1\pm0.3)\times 10^{-5}$
\citep{mikej06}. Assuming that the GRXE spectrum is a power-law with
photon index $\Gamma=2$, we can rescale this factor to the 1--7 keV
band: $F_{\rm 1-7 keV}/F_{\rm 3.5 \mu m}= (4.2\pm0.3)\times
10^{-5}$. Multiplying this ratio by the measured NIR flux yields the
expected GRXE intensity in the considered direction:
$(5.9\pm1.2)\times 10^{-11}$ \flux\ deg$^{-2}$. Consequently, the GRXE
flux from the entire {\sl Chandra} field is expected to be $F_{\rm
1-7~keV} = (8.4\pm1.8) \times 10^{-13}$ \flux. This estimate does not
take into account the significant interstellar absorption in the studied
direction. Considering column densities $N_{\rm H}\sim$ 0.5--4 $\times10^{22}$
cm$^{-2}$ to be typical of Galactic X-ray sources in this
field (see the discussion of this issue in Sect.~\ref{sect:data}
above), the expected absorbed GRXE flux $F_{\rm
1-7~keV,~absorbed}\sim$ (4--7) $\times 10^{-13}$ \flux. 

The above estimate still does not take into account the
contribution of extragalactic sources. The all-sky average intensity
of the cosmic X-ray background (CXB) is $I_{\rm 2-10~keV}\sim 2\times 
10^{-11}$ \flux\ deg$^{-2}$ \citep[e.g.][]{mikej_cxb,mikej_heao1,hickox06}, 
or $I_{\rm 1-7~keV}\sim 1.8\times10^{-11}$ \flux\ deg$^{-2}$ (assuming
a power-law spectrum with $\Gamma=1.4$). This translates into a
flux $F_{\rm 1-7~keV}\sim 2.5\times10^{-13}$ \flux\ for our {\sl
Chandra} field. Taking into account that the full Galactic column
density in the direction of our study is $N_{\rm H}\sim 6\times 10^{22}$
cm$^{-2}$, the CXB contribution to the observed X-ray flux is
expected to be $\sim 1.2\times10^{-13}$ \flux. Therefore, the
total (GRXE plus CXB) X-ray flux from the studied region should be
$F_{\rm 1-7~keV,~expected}\sim$ (5--8) $\times 10^{-13}$ \flux.  
This is compatible with the flux actually measured by {\em
Chandra}: $F_{\rm 1-7~keV,~observed}=(1.0\pm0.1)\times 10^{-12}$ 
\flux. We should stress here that correcting for the interstellar
absorption is very important for the 1--7~keV energy band but is not
straightforward and subject to significant uncertainties due to
possible variations in the absorption along the line of sight and
across the studied field.   

The consistency of cumulative NIR and X-ray characteristics of the
studied {\sl Chandra} field provides additional support to our adopted Galaxy 
model, which is used in the subsequent analysis. We emphasize that it
is crucial that the adopted mass model correctly reproduces the NIR
flux from the considered direction of the GP.  

\subsection{Luminosity function of weak Galactic X-ray sources}

Despite the long history of studying weak ($L_{\rm x}<10^{34-35}$
\lum) Galactic X-ray sources, their luminosity function remained
poorly constrained until lately. We recently used two all-sky X-ray
surveys (the RXTE slew survey in the 3--20~keV band and 
the ROSAT all-sky survey below 2~keV) to construct a luminosity
function of X-ray sources located near the Sun in the 
broad range $10^{27}$--$10^{34}$ \lum\ \citep{sazonov06}. Below we use
this luminosity function in conjunction with the Galaxy model
described above for predicting the number-flux function of sources in the
studied GP region. 

\section{Results}

\subsection{X-ray images}

If there were no significant population of weak Galactic X-ray
sources, one could expect the surface density of point sources
to be similar in different regions of the sky observed by {\it
Chandra} to the same depth\footnote{Although counts of
extragalactic sources exhibit small field-to-field variations due to
the large-scale structure of the Universe (see
e.g. \citealt{voss06,cappelluti07}).}. However, if we compare the {\sl Chandra}
image of the GP region with those of the CDFN and CDFS obtained with
the same exposure (Fig.~\ref{three_images}), we clearly see that 
the surface density of point sources is much higher in the GP region. 

This difference becomes even more pronounced if
we take into account that due to the considerable 
interstellar absorption in the direction of $l\sim28.5^\circ$,
$b\sim0.0^\circ$, the number of detected extragalactic sources in that field
is expected to be significantly reduced compared to the CDFN and 
CDFS. For extragalactic sources with fluxes $10^{-15}$--$10^{-14}$ \flux\ 
in the 1--7 keV energy band, whose spectra can typically be described as power
laws with photon indexes $\Gamma\sim$ 1.4--1.8
\citep[e.g.][]{hickox06}, the Galactic absorption with $N_{\rm H}\sim
6\times10^{22}$ cm$^{-2}$ is expected to suppress the {\it Chandra}
count rates in the 1--7 keV band by a large factor of $\sim$ 4.3--5.3 (note
that since absorption mostly removes photons with energies $\sim$
1--2~keV, the corresponding decrease in the energy flux at 1--7~keV is
much smaller). Therefore, for a given exposure time, {\sl Chandra}
should detect in the CDFN and CDFS roughly 5 times as many
extragalactic sources as in the GP field.

The much higher surface density of sources in the GP region is 
further reflected in the log $N$--log $S$ function of detected sources
(see Fig.~\ref{logn_logs} below). We conclude that in
the Galactic plane weak point X-ray sources of Galactic origin greatly 
outnumber extragalactic ones.

\begin{figure}
\includegraphics[width=\columnwidth,bb=0 80 605 700,clip]{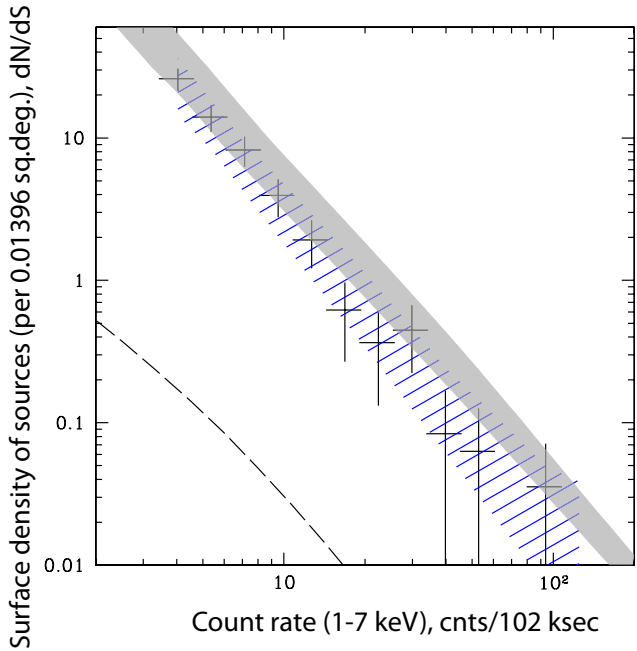}
\caption{Differential number-flux function of sources measured in the
GP region and corrected for the incompleteness at low count rates 
(points with error bars). The hatched region represents the manifold
of number-flux functions allowed by the data (see main text and the
corresponding range of allowed luminosity functions in
Fig.~\ref{nufnu}). The gray band shows the expected range of
number-flux functions based on the luminosity function of weak X-ray
sources in the Solar vicinity from \cite{sazonov06} and including the
predicted contribution of extragalactic sources. The width of this
region is determined by the $1\sigma$ uncertainties in the luminosity
function of local sources. Count rates of Galactic sources were
reduced to take into account the typical interstellar absorption in
the GP region ($N_{\rm H}\sim 2\times 10^{22}$~cm$^{-2}$, see main
text). The expected log $N$--log $S$ distribution of extragalactic
sources (dashed line) was also modified by the Galactic absorption
($N_{\rm H}\sim 6\times 10^{22}$~cm$^{-2}$). One {\sl
Chandra} count of used 102 ksec observation corresponds to unabsorbed fluxes of $2.5\times10^{-16}$ 
\flux\  and $5\times10^{-16}$ \flux\ for Galactic ($N_{\rm H}\sim 2\times
10^{22}$~cm$^{-2}$, $\Gamma=2.0$) and extragalactic ($N_{\rm H}\sim 6\times
10^{22}$~cm$^{-2}$, $\Gamma=1.8$) sources, respectively.}
\label{diff_logn_logs} 
\end{figure}

\begin{figure}
\includegraphics[height=\columnwidth,bb=44 22 600 786,clip,angle=-90]{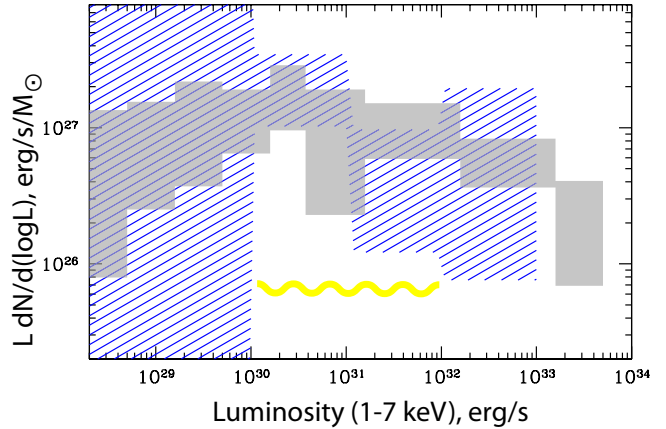}
\caption{Range of luminosity functions of Galactic X-ray sources
in the GP region (hatched area) allowed by the measured number-flux
function  (see Fig.~\ref{diff_logn_logs}). This is compared with the
luminosity function (gray area) of weak X-ray sources in the Solar
neighborhood \citep{sazonov06}. For the latter we included the
uncertainty in the fraction of young single stars in the studied
GP region (relative to cataclysmic variables
and coronally active binary stars, see \citealt{sazonov06} for
details). Specifically, we allowed this contribution to vary from zero
to the value measured near the Sun. Note that we effectively probe
only the luminosity range 
$10^{30}$--$10^{32}$ \lum\ (as indicated by  the wavy line).} 
\label{nufnu}
\end{figure}

\subsection{Log $N$--log $S$ distribution}

We now address the number-flux function of detected sources. In
Fig.~\ref{diff_logn_logs} we show the differential log $N$--log $S$
distribution measured by {\sl Chandra} in the GP region and corrected for the
incompleteness near the detection threshold (see
Sect.~\ref{sect:data}). These data are compared with
the results of our modelling described below. 

To make a prediction for the number-flux function, we convolved the
luminosity function of weak X-ray sources measured in the Solar
vicinity \citep{sazonov06} with the Galactic disk model described in
Sect.~\ref{disk_model}. To convert the 2--10 keV luminosity function
of \cite{sazonov06} into the 1--7~keV band, we assumed a power-law
spectrum with $\Gamma=2.0$. In addition, to take into account the interstellar
absorption in the considered direction, we multipled all predicted
1--7~keV count rates by a factor of 0.34, which corresponds to
a power-law spectrum with $\Gamma=2.0$ absorbed by neutral gas with
column density $N_{\rm H}=2\times10^{22}$ cm$^{-2}$. As we discussed
in Sect.~\ref{sect:data}, this is only a rough (but reasonable)
approximation of the true situation, as in 
reality there is a large scatter in $N_{\rm H}$ values from source to
source. 

As regards the expected contribution of extragalactic sources, we
estimated it from the log $N$--log $S$ curves measured in the CDFN and
CDFS by applying an absorption correction as appropriate for our GP
field ($N_{\rm H}\sim 6\times 10^{22}$~cm$^{-2}$). Adding this expected
extragalatic contribution (shown by the dashed line in
Fig.~\ref{diff_logn_logs}) to the expected contribution of Galactic
sources yields the expected range of log $N$--log $S$ functions, shown
in gray in Fig.~\ref{diff_logn_logs}. There is apparently a good
agreement between the expected and observed distributions. The
apparent small difference between them may well be caused by some
inaccuracy in our correction for the interstellar absorption.

On the other hand, we can use the observed number-flux function to
put constraints on the luminosity function of point sources
in the studied GP region. To this end, we assumed that the luminosity
function in a $L^2 dN/dL$ representation can be described by a set of
constants in the following intervals: $10^{28}$--$10^{29}$,
$10^{29}$--$10^{30}$, $10^{30}$--$10^{31}$, $10^{31}$--$10^{32}$, and
$10^{32}$--$10^{33}$~\lum. By allowing these constants to vary and
convolving the resulting luminosity functions with our  
Galactic disk model, we obtained a manifold of trial number-flux
functions of sources. These were then compared in terms of the reduced
$\chi^2$ value ($\chi^2$ per degree of freedom) with the observed log
$N$--log $S$ function (taking into account the estimated contribution of 
extragalactic sources). Those trial luminosity functions that resulted
in a reduced $\chi^2$ of less than 1.5 were regarded as allowed by the
data. The resulting allowed range of number-flux functions (including
extragalactic sources) is shown as the dashed area in
Fig.~\ref{diff_logn_logs}, and the corresponding range of acceptable luminosity
functions of Galactic sources is shown as the dashed area in
Fig.~\ref{nufnu}, where it is compared with the luminosity function of local
sources taken from \citet{sazonov06}. 

As follows from Fig.~\ref{nufnu}, the allowed range of luminosity
functions in the GP region is fully compatible with the luminosity
function of point sources in the Solar vicinity. It is important to
note here that the $\sim$100~ksec {\it Chandra} observation of the 
GP region allows us to efficiently constrain the luminosity function
only in the range $10^{30}$--$10^{32}$ \lum. The upper bound appears
due to the limited stellar mass within the small solid angle covered
by {\it Chandra} ($< {\rm few}\times10^6 M_\odot$). Sources brighter
than $\sim 10^{32}$~\lum\ are simply too rare to be found in such a
small area of the Galactic disk. The lower bound is due to finite
{\sl Chandra}'s sensitivity.

\begin{figure}
\includegraphics[width=\columnwidth]{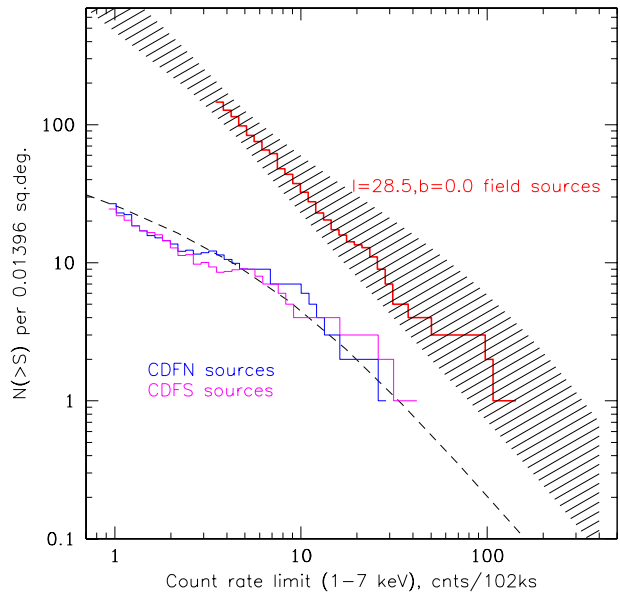}
\caption{Cumulative number-flux functions of point sources detected
in the GP region (upper histogram) and in extragalactic fields ({\sl
Chandra} Deep  Fields North and South). All data are corrected for the
incompleteness at low count rates. The dashed line shows an
approximation of the extragalactic source counts by a simple analytic
model similar to that of \cite{moretti02}. One {\sl
Chandra} count of used 102 ksec observation corresponds to unabsorbed fluxes of $2.5\times10^{-16}$ 
\flux\  and $5\times10^{-16}$ \flux\ for Galactic ($N_{\rm H}\sim 2\times
10^{22}$~cm$^{-2}$, $\Gamma=2.0$) and extragalactic ($N_{\rm H}\sim 6\times
10^{22}$~cm$^{-2}$, $\Gamma=1.8$) sources, respectively.}
\label{logn_logs}
\end{figure}

In Fig.~\ref{logn_logs} we compare the cumulative log
$N$--log $S$ distributions constructed for the GP field and for the CDFS and
CDFN (all corrected for the incompleteness at low count rates). One
can clearly see how Galactic sources become progressively more
dominant toward lower fluxes.

\subsection{Resolved fraction of the GRXE}

Integrating the number-flux function measured in the GP region
(Fig.~\ref{logn_logs}) down to our effective detection limit of 5 counts
(over the 102 ksec exposure, which corresponds to an absorption corrected
flux of $1.2\times 10^{-15}$ \flux\ in the 1--7~keV band for Galactic
sources, assuming $N_{\rm H}=2\times10^{22}$ cm$^{-2}$) yields $\sim 1.6$~kcnts
out of the total $\sim 6.5$~kcnts collected by {\sl Chandra}. This
implies that $\sim 25$\% of the total flux at energies 1--7~keV from the
GP region is already resolved by {\sl Chandra} into point
sources. If we similarly integrate the expected log $N$--log $S$ function of
extragalactic sources (aslo shown in Fig.~\ref{logn_logs}) down to the
same count rate limit (which corresponds to a somewhat higher absorption
corrected flux of $2.5\times10^{-15}$ \flux\ for $N_{\rm H}\sim
6\times10^{22}$~cm$^{-2}$), we find that $\sim 6$\% of the total flux
is resolved into extragalactic sources. 

This implies that at least $\sim 19$\% of the GRXE is due to Galactic
point sources, presumably cataclysmic variables and coronally active
stars. Moreover, an extrapolatation of the measured log $N$--log
$S$ curve to fluxes below the current detection limit based on the
luminosity function of local weak X-ray sources \citep{sazonov06} is
consistent with all of the X-ray flux from the GP being due to point
sources, mostly of Galactic origin. These conclusions are consistent
with our results for the Galactic Center region \citep{mikej_gc}.

\begin{acknowledgements}

This research made use of data obtained from the High Energy Astrophysics
Science Archive Research Center Online Service, provided by the
NASA/Goddard Space Flight Center. We acknowledge the use of the
Legacy Archive for Microwave Background Data Analysis (LAMBDA).
Support for LAMBDA is provided by the NASA Office of Space Science.

\end{acknowledgements}

\end{document}